\documentclass[12pt]{iopart}

\usepackage{iopams}
\usepackage{graphicx}
\begin{document}

\title[]{Photon-echo-based quantum memory for optical squeezed states}

\author{Miao-Xin Wu, Ming-Feng Wang, and Yi-Zhuang Zheng$^{\dag}$}

\address{School of Physics ${\&}$ Electronics Information Engineering, Wenzhou University\rm,  Wenzhou 325027, People's Republic of China}
\ead{yzzheng@wzu.edu.cn}
\begin{abstract}
The ability to efficiently realize storage and readout of optical squeezed states plays a key roll in continuous-variables quantum information processing. Here we study the quantum memory (QM) for squeezed state of propagating light in atoms based on the hybrid photon echo re-phasing (HYPER). The optical quantum state is recorded in two sublevels of the ground state of an atomic ensemble to realize long-lived QM. Taking into account the noise effect due to atomic decay, our estimation indicates that, with currently available technique, high fidelities larger than the classical fidelity threshold 81.5 ${\%}$ are obtainable. Our work provides some practical guidance for realization of efficient and faithful photon-echo-based memory for squeezed light.

\end{abstract}

\maketitle

\section{Introduction}
QM for light plays a key role in quantum-information processing and has been considered as a basic ingredient for quantum repeaters [1] and scalable all-optical quantum computers [2]. Nowadays, much effort has been devoted to such QM through different approaches, including Faraday rotation [3, 4], electromagnetically induced transparency (EIT) [5, 6], off-resonant Raman transitions [7, 8], and photon echoes [9-12]. Photon-echo-based QMs are currently attracting particular attention, not only for their abilities to implement in solid-state materials, but also for their successful achievement in storage efficiency [13, 14] and multimode-memory capacity [15]. By now, in this area, many impressive results have been achieved. Gisin {\it et al.} showed [16] that photon-echo techniques can be used to store and retrieve time-bin qubits [17]. Clausen {\it et al.} demonstrated [18] the storage of entangled photons using atomic frequency comb (AFC) technique. Recently, high efficient storage of optical coherent state (about 87${\%}$) based on gradient echo memory (GEM) has also been demonstrated [14]. Despite these achievements, however, optical squeezed-state storage based on photon-echo techniques, as far as we know, has not yet been studied both theoretically and experimentally.

Squeezed state of light (SSL), which provides a reduction in one of its quadratures below the standard quantum limit, plays a fundamental role in quantum-information processing. More specifically, a QM for SSL has been shown to be especially valuable for interferometry [19], high-precision measurement [20], and quantum information with continuous variables [21]. During the last decade, a variety of approaches concerning the storage of SSL have been proposed. One of them proposes to transfer the squeezed state from non-classical light to the excited atomic states [22]. Another approach to map the SSL onto atomic ensemble based on stimulated Raman scattering has been developed by Kozhekin {\it et al.} [8]. In particular, storage and retrieval of SSL with EIT was also investigated [6]. Very recently, Jensen {\it et al.} demonstrated [4] the QM for entangled two-mode squeezed states via Faraday rotation.

In this paper, we study the long-lived QM for SSL based on HYPER---a variant of the two-pulse photon echo (2PE) technique [23]. 2PE is well known to be unsuitable for serving as a QM technique, because of the irritating collective emission of photons (amplified spontaneous emission). In order to overcome this problem, McAuslan {\it et al.} developed a new photon-echo technique called HYPER, which uses electrical field gradients to eliminate the unwanted gain created by the strong re-phasing pulse ($\pi$-pulse [23] or frequency-chirped pulse [24]). Such technique has the outstanding characteristic of avoiding spectral hole-burning, which enable us to implement large-bandwidth and high-efficiency QMs. Here, utilizing this approach, we detailedly derived the processes for storage and retrieval of SSL. We find that, for the ideal case, the squeezed-quadrature variance (SQV) of the retrieved optical field relies only on the optical depth of atomic ensemble. The larger the optical depth, the more squeezing of the input mode will be preserved. Taking into account the noise effect due to atoms coupled to thermal reservoirs, the SQV now depends not only on the optical depth but also on the decay rate. We show that substantial squeezing of the input mode can also be preserved even with currently available technique [23]. The rest of this paper is organized as follows. In Sec. II, we present the basic equations governing the evolution of the quantized atomic and photonic fields based on photon-echo techniques. Next, storage and retrieval of SSL is considered. In Sec. IV, we discuss the possibility of implementing the technique in the solid-state materials. Section V gives our conclusions.

\section{The storage process}
\subsection{The basic evolutions}
\indent ~~~~Our storage material consists of a large number of atoms with an excited state ${\left| e \right\rangle}$ and two lower states ${\left| g \right\rangle}$, ${\left| s \right\rangle}$, see Fig. 1(a). Let us first study the absorption process. Consider a quantized optical pulse being injected into the storage medium to couple ${\left| g \right\rangle}$ and  ${\left| e \right\rangle}$. In the one-dimensional light propagation model, the negative frequency part of the optical field can be decomposed in forward and backward modes
\begin{equation}              
\hat a\left( {z,t} \right) = \varepsilon {\hat a_f}\left( {z,t} \right){e^{ - i\left( {{\omega _0}t - kz} \right)}} + \varepsilon {\hat a_b}\left( {z,t} \right){e^{ - i\left( {{\omega _0}t + kz} \right)}}\;,
\end{equation}
where ${\varepsilon  = \sqrt{ \hbar \omega _{0}/\left( \varepsilon _{0}V \right)}}$, ${\omega _{0}}$ is the central frequency of optical pulse, ${\varepsilon _0}$ is the vacuum permeability, and ${V}$ is the quantized volume. ${\hat a_{k}\left( {z,t} \right)}$ (where ${k = f, b}$) represents the optical operators with the commutation relation${\left[ {{{\hat a}_k}\left( {z,t} \right),\hat a_k^\dag \left( {z,t'} \right)} \right] = \delta \left( {t - t'} \right)}$. For atomic operators, the mean field per atom is defined as [25]
\begin{equation}              
{\hat \sigma _{ge}}\left( {z,t,\Delta } \right) = \frac{1}{{N\left( {\Delta ,z} \right)}}\sum\limits_{n = 1}^{N\left( {\Delta ,z} \right)} {{{\left| g \right\rangle }_{nn}}\left\langle e \right|}\;,
\end{equation}
where ${\Delta = \omega_{eg} - \omega_{0}}$ is the detuning from resonance, with the atomic transition frequency  ${\omega_{eg}}$. In the above sum, the index ${n}$ runs over all atoms ${N\left( {\Delta ,z} \right):= \rho \delta zg\left( \Delta  \right)\delta \Delta }$ within the infinitesimal slice ${\delta z}$ and ${\delta \Delta}$. The notation ${\rho}$ denotes the number density of atoms, and ${g (\Delta)}$ is the normalized atomic spectral distribution. In analogy with the optical field, the negative-frequency part of the atomic operator can be described by the two counter-propagating contributions
\begin{equation}              
{\hat \sigma _{ge}}\left( {z,t,\Delta } \right) = {\hat \sigma _f}\left( {z,t,\Delta } \right){e^{ - i\left( {{\omega _0}t - kz} \right)}} + {\hat \sigma _b}\left( {z,t,\Delta } \right){e^{ - i\left( {{\omega _0}t + kz} \right)}}\;.
\end{equation}
In the low-excitation limit, if the atoms are initially prepared in the ground state, the operators associated to the atomic coherence ${\hat \sigma _k}$ have the following commutation relation [26]
\begin{equation}              
\left[ {{{\hat \sigma }_k}\left( {z,t,\Delta } \right),\hat \sigma _k^\dag \left( {z',t,\Delta '} \right)} \right] = \frac{1}{{\rho g\left( \Delta  \right)}}\delta \left( {z - z'} \right)\delta \left( {\Delta  - \Delta '} \right)\;.
\end{equation}
\indent ~~~~With these definitions, the dynamic of the optical absorbing process can be described by the following equations
\begin{equation}              
\left( {\frac{1}{c}{\partial _t} + {\partial _z}} \right){\hat a_f}\left( {z,t} \right) = ig\rho \int_{ - \infty }^\infty  {d\Delta } g\left( \Delta  \right){\hat \sigma _f}\left( {z,t,\Delta } \right)\;,
\end{equation}
\begin{equation}              
{\partial _t}{\hat \sigma _f}\left( {z,t,\Delta } \right) = \left( {i\Delta  - \frac{\Gamma }{2}} \right){\hat \sigma _f}\left( {z,t,\Delta } \right) + ig{\hat a_f}\left( {z,t} \right) + {\hat f_1}\left( {z,t,\Delta } \right)\;,
\end{equation}
where ${g}$ is the coupling constant of the light-atoms interactions, and we have considered the decoherence of atomic system with ${\Gamma }$ the decay rate of the excited state. ${{\hat f_1}\left( {z,t,\Delta } \right)}$ is the Langevin noise operator, satisfying the following correlation functions [27]
\begin{equation}              
\left\langle {{{\hat f}_1}\left( {z,t,\Delta } \right)\hat f_1^\dag \left( {z',t',\Delta '} \right)} \right\rangle  = \frac{\Gamma }{{\rho g\left( \Delta  \right)}}\delta \left( {z - z'} \right)\delta \left( {t - t'} \right)\delta \left( {\Delta  - \Delta '} \right)\;,
\end{equation}
\begin{equation}              
\left\langle {\hat f_1^\dag \left( {z',t',\Delta '} \right){{\hat f}_1}\left( {z,t,\Delta } \right)} \right\rangle  = 0\;.
\end{equation}
\indent ~~~~The solution of Eq. (6) can easily be found to be
\begin{eqnarray}             
{\hat \sigma _f}\left( {z,t,\Delta } \right) & = & {\hat \sigma _f}\left( {z,{t_0},\Delta } \right){e^{\left( {i\Delta  - \Gamma /2} \right)\left( {t - {t_0}} \right)}} + ig\int_{{t_0}}^t {ds} {\hat a_f}\left( {z,s} \right){e^{\left( {i\Delta  - \Gamma /2} \right)\left( {t - s} \right)}}
 \nonumber \\
 && + \int_{{t_0}}^t {ds} {\hat f_1}\left( {z,s,\Delta } \right){e^{\left( {i\Delta  - \Gamma /2} \right)\left( {t - s} \right)}}\;,
\end{eqnarray}
where ${\hat \sigma _{f}\left( {z,{t_0},\Delta } \right)}$ is the initial atomic operator. To derive the time evolution of optical field, we insert Eq. (9) into Eq. (5) to generate
\begin{eqnarray}             
{\partial _z}{{\hat a}_f}\left( {z,t} \right) &=&  - {g^2}\rho \int_{{t_0}}^t {ds} {{\hat a}_f}\left( {z,s} \right){e^{ - \Gamma \left( {t - s} \right)/2}}\int_{ - \infty }^\infty  {d\Delta } g\left( \Delta  \right){e^{i\Delta \left( {t - s} \right)}}
\nonumber\\&& + ig\rho \int_{ - \infty }^\infty  {d\Delta } g\left( \Delta  \right)\times {{\hat \sigma }_f}\left( {z,{t_0},\Delta } \right){e^{\left( {i\Delta  - \Gamma /2} \right)\left( {t - {t_0}} \right)}}\nonumber\\&& + ig\rho \int_{ - \infty }^\infty  {d\Delta } g\left( \Delta  \right)\int_{{t_0}}^t {ds} {{\hat f}_1}\left( {z,s,\Delta } \right){e^{\left( {i\Delta  - \Gamma /2} \right)\left( {t - s} \right)}}\;.
 \end{eqnarray}
Here, we have neglected the temporal derivative in Eq. (5), since we consider the regime ${\tau  \gg L/c}$ (${\tau}$ is the temporal length of the incident optical pulse; ${L}$ is the length of medium), which allows us to ignore the temporal retardation effect of light. In order to find the solution of Eq. (10), we assume the atomic spectral distribution is uniform, that is ${g\left( \Delta  \right) \sim 1/\gamma }$ (where ${\gamma }$  is the bandwidth of the ensemble), which is justified since, for natural inhomogeneous broadening systems, such as rare-earth-ion doped crystals, the ratio between inhomogeneous line-width and homogeneous line-width can be approximately achieved 6 or more orders of magnitude [28]. Furthermore, the duration of the absorption process is essentially governed by the temporal length of the signal optical pulse, which ensures the values of ${t - s}$ of order ${\tau}$. Hence, in the limit ${\gamma \tau  \gg 1}$ the Fourier transform  ${\int_{ - \infty }^\infty  {d\Delta } g\left( \Delta  \right){e^{i\Delta \left( {t - s} \right)}}}$ in Eq. (10) will act like a ${\delta}$ function. With these assumptions, Eq. (10) can be directly solved to give
\begin{eqnarray}            
{{\hat a}_f}\left( {z,t} \right) = {{\hat a}_f}\left( {0,t} \right){e^{ - \alpha z/2}}
\nonumber\\+ ig\rho \int_0^z {dz'} \int_{ - \infty }^\infty  {d\Delta } g\left( \Delta  \right){{\hat \sigma }_f}\left( {z',{t_0},\Delta } \right){e^{\left( {i\Delta  - \Gamma /2} \right)\left( {t - {t_0}} \right)}}{e^{\alpha \left( {z' - z} \right)/2}}
\nonumber\\+ ig\rho \int_0^z {dz'} \int_{ - \infty }^\infty  {d\Delta } g\left( \Delta  \right)\int_{{t_0}}^t {ds} {{\hat f}_1}\left( {z,s,\Delta } \right){e^{\left( {i\Delta  - \Gamma /2} \right)\left( {t - s} \right)}}{e^{\alpha \left( {z' - z} \right)/2}}\;,\nonumber\\
\end{eqnarray}

\noindent where ${\hat a_{f}\left( {0,t} \right)}$ is the input photonic field, and we have defined the absorption coefficient ${\alpha  = 2\pi {g^2}\rho /\gamma }$. The first term describes the exponential decay of the incoming optical pulse.

\subsection{Long-lived memory for SSL}
\begin{figure}[htb]
\begin{center}
  \includegraphics[width=0.5\textwidth]{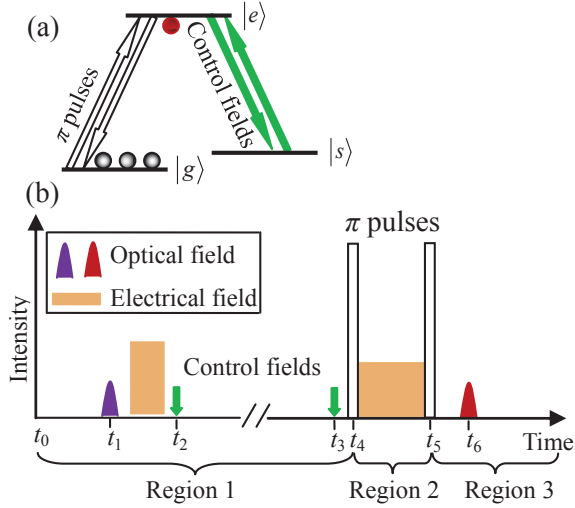}
  \caption{(Color online) (a) Atomic-level structure of storage medium. (b) Pulse sequence of the long-lived HYPER protocol. In region 1, the optical data is absorbed by the storage medium at time ${t_1}$. Subsequently, a pair of control fields (${\pi}$ pulses or frequency-chirped pulses) are applied to the atoms to transform the atomic coherence ${\hat \sigma _{ge}}\rightarrow \hat \sigma _{gs}$ and ${\hat \sigma _{gs}}\rightarrow \hat \sigma _{ge}$. Region 2 realizes the rephasing of the stored light, and finally in region 3 the optical pulse comes out.}
\end{center}
\end{figure}

Now, we consider a long-time storage protocol based on HYPER technique. This protocol consists of three separate time regions, as shown in Fig. 1(b). In region 1 ${(t_{0} < t < t_{4})}$, just after absorbing the signal optical field, a linear electrical field gradient is applied to the sample to cause an additional atomic phase change depending on the position of the atoms. Soon after this field, a strong control optical field coupling to ${\left| e \right\rangle}$ , ${\left| s \right\rangle}$  is turned on to transfer the population from ${\left| e \right\rangle}$ to ${\left| s \right\rangle}$. Obviously, information about signal field is now stored in the spin wave. After a time interval ${(t_{3} - t_{2})}$, another strong optical field propagating in the same direction are applied to drive the population back onto the excited state. In region 2 ${(t_{4} < t < t_{5})}$, two ${\pi}$ pulses sandwiched with an electrical field are applied to the atoms to rephase the input optical mode. In region 3 ${(t > t_{5})}$, the ensemble will emits an expected echo at ${t_{6}}$.

During the storage process, for simplicity, we assume that the first electrical field is short enough, which enables us to ignore the decay of the excited state during the time interval ${(t_{1} < t < t_{2})}$. We also neglect spontaneous decay of the state ${\left| s \right\rangle}$, which is reasonable since we assume the time interval ${(t_{3} - t_{2})}$ is much smaller than the lifetime of ${\left| s \right\rangle}$. Furthermore, the time spent in the region 2 is also temporally short, which allows us to reduce the dynamics of this region to be an instantaneous operation at ${t_{5}}$ [23]. Besides, if the two ¦Ð pulses used in region 2 propagate in opposite directions, the system will satisfy the condition of the backward retrieval resulting in the capability for achieving 100 ${\%}$ QM efficiency [25]. After applying the second ¦Ð pulse in region 2, the atoms start to rephase. At ${t_{6} = 2 t_{5} - 2 t_{4} + t_{3} - t_{2} + t_{1}}$, an echo propagating along negative ${z}$-axis through the medium occurs, which can be described by the following equations

\begin{equation}              
 - {\partial _z}{\hat a_b}\left( {z,t} \right) = ig\rho \int_{ - \infty }^\infty  {d\Delta } g\left( \Delta  \right){\hat \sigma _b}\left( {z,t,\Delta } \right)\;,
\end{equation}
\begin{equation}              
{\partial _t}{\hat \sigma _b}\left( {z,t,\Delta } \right) = \left( {i\Delta  - \frac{\Gamma }{2}} \right){\hat \sigma _b}\left( {z,t,\Delta } \right) + ig{\hat a_b}\left( {z,t} \right) + {\hat f_2}\left( {z,t,\Delta } \right)\;,
\end{equation}
where ${\hat f_{2}\left( {z,t,\Delta } \right)}$ is the Langevin noise of the retrieval process. In deriving the above equations, we have used the boundary condition of the atomic field created at ${t_{5}}$
\begin{equation}              
{\hat \sigma _b}\left( {z,t = 0;\Delta } \right) = {\hat \sigma _f}\left( {z,t = 0;\Delta } \right){e^{ - i\Delta T}}\;,
\end{equation}
where we have set ${t_{5} = 0}$, leading to the QM time ${T = 0 - t_{1}}$. In accordance with (9), (12), and (14), one can easily derive the output optical field
\begin{eqnarray}             
{{\hat a}_b}\left( {0,t} \right) = {{\hat a}_b}\left( {L,t} \right){e^{ - \alpha L/2}} + \alpha {e^{ - \Gamma T/2}}\int_L^0 {dz} {{\hat a}_f}\left( {z,t - T} \right){e^{ - \alpha z/2}}
 \nonumber\\- ig\rho \int_L^0 {dz} \int_{ - \infty }^\infty  {d\Delta } g\left( \Delta  \right){{\hat \sigma }_f}\left( {z,{t_0},\Delta } \right){e^{\left( {i\Delta  - \Gamma /2} \right)\left( {t - {t_0}} \right)}}{e^{ - i\Delta T}}{e^{ - \alpha z/2}}
\nonumber\\- ig\rho \int_L^0 {dz} \int_{ - \infty }^\infty  {d\Delta } g\left( \Delta  \right)\int_{{t_0}}^0 {ds} {{\hat f}_1}\left( {z,s;\Delta } \right){e^{\left( {i\Delta  - \Gamma /2} \right)\left( {t - s} \right)}}{e^{ - i\Delta T}}{e^{ - \alpha z/2}}
 \nonumber\\- ig\rho \int_L^0 {dz} \int_{ - \infty }^\infty  {d\Delta } g\left( \Delta  \right)\int_0^t {ds} {{\hat f}_2}\left( {z,s;\Delta } \right){e^{\left( {i\Delta  - \Gamma /2} \right)\left( {t - s} \right)}}{e^{ - \alpha z/2}}\;,
\end{eqnarray}
where ${\hat a_{b}\left( {L,t} \right)}$ is the vacuum input of the backward retrieval optical mode, and we have used the fact that the optical field is out at ${z = 0}$. Next, Eq. (11) is substituted into Eq. (15), and after interchanging the order of the associated double integral ${\int_0^L {dz} \int_0^Z {dz'}  \to \int_0^L {dz'} \int_{z'}^L {dz} }$, we will obtain
\begin{eqnarray}             
{{\hat a}_b}\left( t \right) = {{\hat a}_b}\left( {L,t} \right){e^{ - \alpha L/2}} + {e^{ - \Gamma T/2}}\left( {{e^{ - \alpha L}} - 1} \right){{\hat a}_f}\left( {0,t - T} \right)
\nonumber\\ + ig\rho \int_0^L {dz} \int_{ - \infty }^\infty  {d\Delta } g\left( \Delta  \right){{\hat \sigma }_f}\left( {z,{t_0},\Delta } \right){e^{\left( {i\Delta  - \Gamma /2} \right)\left( {t - {t_0}} \right)}}{e^{ - i\Delta T}}{e^{\alpha z/2}}{e^{ - \alpha L}}
\nonumber\\ + ig\rho \int_0^L {dz} \int_{ - \infty }^\infty  {d\Delta } g\left( \Delta  \right)\int_{{t_0}}^{t - T} {ds} {{\hat f}_1}\left( {z,s;\Delta } \right){e^{\left( {i\Delta  - \Gamma /2} \right)\left( {t - s} \right)}}{e^{ - i\Delta T}}{e^{\alpha z/2}}{e^{ - \alpha L}}
\nonumber\\+ ig\rho \int_0^L {dz} \int_{ - \infty }^\infty  {d\Delta } g\left( \Delta  \right)\int_{t - T}^0 {ds} {{\hat f}_1}\left( {z,s;\Delta } \right){e^{\left( {i\Delta  - \Gamma /2} \right)\left( {t - s} \right)}}{e^{ - i\Delta T}}{e^{ - \alpha z/2}}
\nonumber\\+ ig\rho \int_0^L {dz} \int_{ - \infty }^\infty  {d\Delta } g\left( \Delta  \right)\int_0^t {ds} {{\hat f}_2}\left( {z,s;\Delta } \right){e^{\left( {i\Delta  - \Gamma /2} \right)\left( {t - s} \right)}}{e^{ - \alpha z/2}}\;,
\end{eqnarray}

\noindent where ${t - T}$ is the time when the incident pulse is completely absorbed. The second term shows that information about the input optical mode is now printed onto the output mode. The last three noises terms are independent of each other, since they correspond to the Langevin noises at different times. Integrating Eq. (16) over times of the duration of the input pulse, one obtains the dimensionless optical operator
\begin{eqnarray}             
{{\hat a}_b} = {\tau ^{ - 1/2}}\int_0^\tau  {dt} {{\hat a}_b}\left( t \right)
\nonumber\\={\tau ^{ - 1/2}}{e^{ - \alpha L/2}}\int_0^\tau  {dt} {{\hat a}_b}\left( {L,t} \right)
 \nonumber\\+ {\tau ^{ - 1/2}}{e^{ - \Gamma {t_d}/2}}\left( {{e^{ - \alpha L}} - 1} \right)\int_0^\tau  {dt} {{\hat a}_f}\left( {0,t - T} \right)
 \nonumber\\+ i\sqrt \alpha  \int_0^L {dz} \left\{ {\left[ {\hat D\left( z \right) + {{\hat F}_{11}}\left( z \right)} \right]{e^{\alpha z/2}}{e^{ - \alpha L}} + \left[ {{{\hat F}_{12}}\left( z \right) + {{\hat F}_2}\left( z \right)} \right]{e^{ - \alpha z/2}}} \right\}\;,\nonumber\\
\end{eqnarray}
where we defined the new operators for the initial atomic operator and Langevin noise operators
\begin{eqnarray}
\hat D\left( z \right) = \sqrt {\frac{{\rho \gamma }}{{2\pi \tau }}} \int_0^\tau  {dt} \int_{ - \infty }^\infty  {d\Delta } g\left( \Delta  \right){\hat \sigma _f}\left( {z,{t_0},\Delta } \right){e^{\left( {i\Delta  - \Gamma /2} \right)\left( {t - {t_0}} \right)}}{e^{ - i\Delta T}},\nonumber\\
{\hat F_{11}}\left( z \right) = \sqrt {\frac{{\rho \gamma }}{{2\pi \tau }}} \int_0^\tau  {dt} \int_{ - \infty }^\infty  {d\Delta } g\left( \Delta  \right)\int_{{t_0}}^{t - T} {ds} {\hat f_1}\left( {z,s;\Delta } \right){e^{\left( {i\Delta  - \Gamma /2} \right)\left( {t - s} \right)}}{e^{ - i\Delta T}},\nonumber\\
{\hat F_{12}}\left( z \right) = \sqrt {\frac{{\rho \gamma }}{{2\pi \tau }}} \int_0^\tau  {dt} \int_{ - \infty }^\infty  {d\Delta } g\left( \Delta  \right)\int_{t - T}^0 {ds} {\hat f_1}\left( {z,s;\Delta } \right){e^{\left( {i\Delta  - \Gamma /2} \right)\left( {t - s} \right)}}{e^{ - i\Delta T}},\nonumber\\
{\hat F_2}\left( z \right) = \sqrt {\frac{{\rho \gamma }}{{2\pi \tau }}} \int_0^\tau  {dt} \int_{ - \infty }^\infty  {d\Delta } g\left( \Delta  \right)\int_0^t {ds} {\hat f_2}\left( {z,s;\Delta } \right){e^{\left( {i\Delta  - \Gamma /2} \right)\left( {t - s} \right)}}.\nonumber
\end{eqnarray}
According to Eqs. (4), (8) and (9), the above operators obey
\begin{equation}              
\left\langle {\hat D\left( z \right){{\hat D}^\dag }\left( {z'} \right)} \right\rangle  = A{e^{\Gamma {t_0}}}\delta \left( {z - z'} \right)\;,
\end{equation}
\begin{equation}              
\left\langle {{{\hat F}_{11}}\left( z \right)\hat F_{11}^\dag \left( {z'} \right)} \right\rangle  = \left( {{e^{ - \chi }} - A{e^{\Gamma {t_0}}}} \right)\delta \left( {z - z'} \right)\;,
\end{equation}
\begin{equation}              
\left\langle {{{\hat F}_{12}}\left( z \right)\hat F_{12}^\dag \left( {z'} \right)} \right\rangle  = \left( {A - {e^{ - \chi }}} \right)\delta \left( {z - z'} \right)\;,
\end{equation}
\begin{equation}              
\left\langle {{{\hat F}_2}\left( z \right)\hat F_2^\dag \left( {z'} \right)} \right\rangle  = \left( {1 - A} \right)\delta \left( {z - z'} \right)\;.
\end{equation}
Here we have defined the dimensionless factor ${A = \frac{1}{{\Gamma \tau }}\left( {1 - {e^{ - \Gamma \tau }}} \right)}$ and ${\chi  = \Gamma T}$. Eq. (17) is the main result of this paper. Obviously, in contrast to the input light state the output one is severely distorted. In order to measure how well the output state compares to the input one, we introduce the fidelity ${F = \left\langle \varphi  \right.\left| {{{\hat \rho }_{out}}} \right.\left| \varphi  \right\rangle}$ for light, where ${\left| \varphi  \right\rangle}$ is the input state, and ${\hat \rho _{out}}$ is the output state. Note that the atomic and optical states involved here are all Gaussian. Furthermore, we assume that the input squeezed state is with zero displacement. For such states, the fidelity is given by ${F = \frac{1}{2}\sqrt {\det {{\left( {M + M'} \right)}^{ - 1}}}}$ [29] where ${M}$ and ${M'}$ stand for the covariance matrix of the input and output states, respectively. In accordance with Eqs. (17) - (21), one may directly derive the covariance matrix, and we finally get
\begin{figure}[htbp]
\begin{center}
  \includegraphics[width=0.5\textwidth]{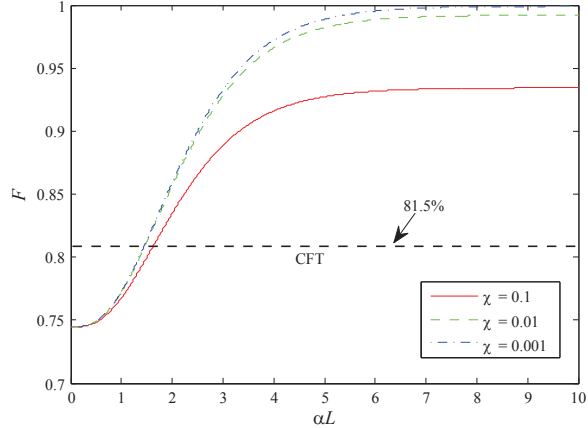}
  \caption{(Color online) Storage fidelities versus optical depth in the presence of atomic decay for 7 dB squeezing input.}
\end{center}
\end{figure}

\vspace{6pt}
\centerline{${F = \frac{2}{{\sqrt {\left[ {\left( {1 + {e^{ - 2r}}} \right) + {e^{ - \chi }}{{\left( {1 - {e^{ - \alpha L}}} \right)}^2}\left( {{e^{ - 2r}} - 1} \right)} \right]\left[ {\left( {1 + {e^{2r}}} \right) + {e^{ - \chi }}{{\left( {1 - {e^{ - \alpha L}}} \right)}^2}\left( {{e^{2r}} - 1} \right)} \right]} }}}$.}
\vspace{6pt}

\noindent In deriving Eq. (22), we have used the fact that ${\left\langle {{{\hat X}_f}\left( {t - {t_d}} \right){{\hat X}_f}\left( {t' - {t_d}} \right)} \right\rangle  = {e^{ - 2r}}\delta \left( {t - t'} \right)}$\\${/4}$, which means that the ${x}$-quadrature of the input optical mode are initially squeezed with the squeezing parameter ${r}$. Note that, for the ideal case ${\chi = 0}$ and the certain squeezing input, the fidelity depends only on the optical depth.  The larger the optical depth, the higher the fidelity. For the case of ${\chi \neq 0}$, in Fig. 2 we show the fidelity in its dependence on the optical depth for different decay parameter ${\chi}$, where the dashed line denotes classical fidelity threshold (CFT) [30]. One can see from the figure that, for optical depth large than 2, the fidelity is well above CFT, and the maximal fidelity is mainly limited by the atomic decay. Fidelity vs optical depth for different squeezing is also depicted in Fig. 3, showing that the more highly squeezed the input mode, the less we can tolerate the losses of the memory process. It should be noted that, for low-squeezed light, its fidelity is much higher than CFT even within small optical depth, which reflects that the quantum features of low-squeezed states are much more harder to be reconstructed via a ¡®¡®measure-and-prepare¡¯¡¯ strategy [30].

\begin{figure}[htbp]
\begin{center}
  \includegraphics[width=0.5\textwidth]{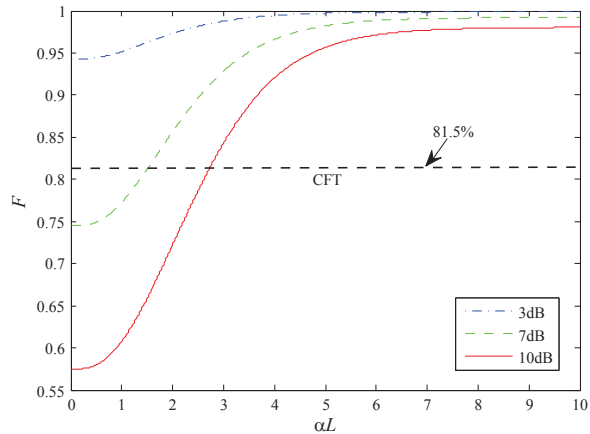}
  \caption{(Color online) The memory fidelities as a function of optical depth for ${\chi}$ = 0.01.}
\end{center}
\end{figure}

\section{Implementation}
We now focus on a possible experimental implementation for the present protocol. We propose to use praseodymium ions doped into an Y${_2}$SiO${_5}$ crystal. This system has been used for implementing an efficient frequency-selective population transfer (qubit distillation) [31] and a long-lived (storage time above 40s) EIT-based memory [32]. In particular, in such system, high efficient photon-echo-based QMs [13] have also been demonstrated. We consider the ${^3}$H${_4}$ to ${^1}$D${_2}$ transition (605.977nm) for site 1 in a 0.02 ${\%}$ Pr${^3}$${^+}$:Y${_2}$SiO${_5}$ crystal [28]. Its homogeneous and inhomogeneous line-width are about ${\gamma}$ = 4.4 GHz and ${\Gamma}$ = 1 KHz, respectively. For light, we suggest preparing the input squeezed optical pulse with duration ${\tau}$ = ${1~\mu s}$, which ensures ${\gamma \tau  \gg 1}$. If we set the storage time ${T = 10~\mu s}$, the value of ${\chi}$ can be determined to be 0.01. For experimental feasible parameter ${\alpha L = 2.5}$~\cite{HYPER}, one can achieve ${F}$ = 98.16 ${\%}$, 89.65 ${\%}$, and 78.67 ${\%}$ for 3 dB, 7 dB, and 10 dB squeezing input, respectively.

\section{Conclusion}
We have studied the long-lived QM for squeezed state of propagating light in a solid-state system based on HYPER. We found that, the HYPER technique has a good performance in storage of SSL. Just as the storage of classical field, the limitation of the QM for SSL mainly comes from the optical depth. The larger the optical depth, the more squeezing of the input mode will be preserved. Besides, the atomic decay is another factor that limits the achievable fidelities, while we show that its influence on the storage process can be suppressed via sacrificing the storage time. In conclusion, our work provides hopeful possibilities for the practical realization of efficient QM for SSL based on photon echoes.

\section{Acknowledgments}
We acknowledge support from the Natural Science Foundation of Zhejiang province (Grants No. LY12A05001), the department of education of Zhejiang province (Y201120838)£¬ and the Natural Science Foundation of China (Grant No. 11074190).

\end{document}